\documentclass{SCGE}
\usepackage{multicol}
\usepackage{multirow}
\usepackage{longtable}
\usepackage{epsfig}

\makeatletter

\newenvironment{figurehere}
  {\def\@captype{figure}}
  {}
\makeatother

\begin{document}

\begin{picture}(0,0){\rm
\put(0,-39){\makebox[160truemm][l]{\bf {\sanhao\raisebox{2pt}{.}}
Research Paper  {\sanhao\raisebox{1.5pt}{.}}}}}
\end{picture}

\def\bm{\boldsymbol}

\def\dl{\displaystyle}
\def\du{\end{document}}

\Year{2010} %
\Month{October}
\Vol{0} %
\No{0} %
\BeginPage{1} %
\EndPage{4} %
\AuthorMark{{\rm QIAN Lei,} }
\DOI{} 

\title{Constraining Photon Mass by Energy-Dependent Gravitational Light Bending}

\author[1]{QIAN Lei}{}

\address[{\rm1}]{National Astronomical Observatories, Chinese Academy of Sciences, Beijing 100012, China;}

\maketitle \vspace{-3.5mm}{\footnotesize\begin{center} 
\end{center}}\vspace*{-5mm}

\begin{center}
\rule{16.5cm}{0.4pt}
\parbox{16.5cm}
{\begin{abstract}
In the standard model of particle physics, photons are mass-less particles with a particular dispersion relation. Tests of this claim at different scales are both interesting and important. Experiments in territory labs and several exterritorial tests have put some upper limits on photon mass, e.g. torsion balance experiment in the lab shows that photon mass should be smaller than $1.2\times 10^{-51}\rm g$. In this work, this claim is tested at a cosmological scale by looking at strong gravitational lensing data available and an upper limit of  $8.71\times 10^{-39}$g on photon mass was given. Observations of energy-dependent gravitational lensing with not yet available higher accuracy astrometry instruments may constrain photon mass better.
\end{abstract}}
\end{center}\vspace*{-0.6cm}
\begin{center}
\parbox{16.5cm}{\bf\jiuhao astrometry, sun, galaxy: nuclei
}
\end{center}

\begin{center}
\parbox{16.5cm}{\PACS{\hspace*{-2mm}\rm 97.80.Fk, 97.10.Cv, 97.60.Bw, 97.20.Rp}
\rule{16.5cm}{0.4pt}}\end{center}



\wuhao\vspace*{1.5mm}
\begin{multicols}{2}
\renewcommand{\baselinestretch}{1.08} \baselineskip 12.2pt\parindent=10.8pt

\no 

\section{Introduction}
Photons are particles mediating electromagnetic force. In the
standard model of particle physics, they are zero-mass particles
with a particular dispersion relation
\begin{equation}
E^2=p^2 c^2,
\label{dispersion}
\end{equation}
where $E$, $p$, and $c$  are energy, momentum, and the speed of light,
respectively. This property is closely related to the inverse square
law of electrostatic force between two charges
\begin{equation}
F=\frac{1}{4\pi \epsilon_0}\frac{q_1 q_2}{r^2},
\label{force}
\end{equation}
where $F$ is the electric force between the two charges $q_1$ and
 $q_2$ with a separation of $r$, and $\epsilon_0$ is the vacuum permittivity.

If a photon has a non-zero mass $m$, the dispersion relation becomes
\begin{equation}
E^2=p^2 c^2+m^2 c^4.
\label{dispersionwithmass}
\end{equation}
Also, the electrostatic potential takes on a Yukawa form of
\begin{equation}
V=\frac{1}{4\pi \epsilon_0}\frac{q}{r}e^{-\mu r},
\label{potential}
\end{equation}
accounting for the finite range of the force mediated by a non-zero
mass particle. Electromagnetism with non-zero mass photon is also
different, and can be described by[1]
\begin{eqnarray}
\nabla\cdot {\bf E}&=&4\pi\rho-\mu^2 V,\label{Maxwell1}\\
\nabla\times {\bf E}&=&-\frac{1}{c}\frac{\partial {\bf B}}{\partial t},\label{Maxwell2}\\
\nabla\cdot {\bf B}&=&0,\label{Maxwell3}\\
\nabla\times {\bf B}&=&\frac{4\pi}{c}{\bf J}+\frac{1}{c}\frac{\partial {\bf E}}{\partial t}-\mu^2{\bf A},\label{Maxwell4}
\end{eqnarray}
where ${\bf E}$, ${\bf B}$, ${\rho}$, ${\bf J}$, $V$, ${\bf A}$  and
$c$ are electric field, magnetic induction,
charge density, current density, scalar potential,
vector potential and the speed of light, respectively, while
$\mu^{-1}=\hbar/mc$ is the Compton wavelength of photons, with
$m$  denoting photon mass. Note that in equation~(\ref{Maxwell1}) and equation~(\ref{Maxwell4}), 
there are additional terms $\mu^2 V$ and $\mu^2 {\bf A}$,
which vanish when photon mass $m$ is zero, and then the above
equations are just the Maxwell equations.

These effects caused by a non-zero photon mass can be used to
constrain photon mass itself. The dispersion relation
(\ref{dispersionwithmass}) means that non-zero mass photons with
higher energies travel faster than lower energies ones. Therefore, 
measurements of the difference in arrival times of photons with
different energies emitted simultaneously at the same place can
be used to constrain photon mass. This has been done by
analyzing gamma-ray burst data[2]. The key assumption is that
photons of different energies were emitted simultaneously,
however, in reality this is not guaranteed, and this will limit the
accuracy of the final constraint on photon mass.

Deviations from the inverse square law of electrostatic force can also be constrained by torsion balance
experiments in territory labs, which infer an upper limit on photon mass of $1.2\times 10^{-51}\rm g$[3]. This is an effective way to constrain photon mass. However, it is a small scale test, which would not be effective if the Compton length of photons
$\mu^{-1}=\hbar/mc$ is large (say, as large as a galaxy). Experiments have also been done to constrain
the $\mu^2{\bf A}$ term in equation~(\ref{Maxwell4}). With the estimation of $\bf A$, one can get a constraint
on the Compton length of photons (also on photon mass)[1]. This method is limited by the accuracy of the
estimation of ${\bf A}$. Some authors also mentioned the constraint on photon mass from the measurements
of light bending by the sun[4-6], the estimated upper limit is about $10^{-40}\rm g$. For a more detailed review on constraining photon mass, one may refer to e.g. [7].

In this paper, we try to investigate the constraints on photon
mass at a cosmological scale with strong gravitational lensing data.
In section 2, we present the equations used in our analysis. Results
are given in section 3. We then do some discussion in section 4.
Hereafter in this paper we use natural units ($c=\hbar=1$) for
simplicity.

\section{Equations}
The bending of light within a gravitational field can be considered as the scattering of photons
in this gravitational field[4]. The gravitational field is treated classically as a background.

The dispersion relation of a photon of energy $E$ and mass $m$ in the
gravitation field of an object with mass $M$ is
\end{multicols}
\rule{4cm}{1pt}
\begin{equation}
\begin{split}
2\left(\frac{GM}{b}\right)^2=\frac{1-\cos\theta}{\left(2+\frac{m^2}{E^2-m^2}\right)^2
+\frac{8}{3}\left(1+\frac{m^2}{E^2-m^2}\right)(1-\cos\theta)\ln(1-\cos\theta)-\frac{2}{3}(1-\cos\theta)^2},
\end{split}
\label{dispersion2}
\end{equation}
\hspace{13cm} \rule{4cm}{1pt}
\wuhao\vspace*{1.5mm}
\begin{multicols}{2}
\renewcommand{\baselinestretch}{1.08} \baselineskip 12.2pt\parindent=10.8pt
\noindent where $G$ is the gravitational constant, $b$ is the impact
parameter and $\theta$ is the deflection angle. When $\theta$ is
small and in the extreme relativistic limit $E\gg m$, this equation
can be approximated as
\begin{equation}
\theta=\theta_E\left(1+\frac{m^2}{2E^2}\right),
\label{deflection}
\end{equation}
where $\theta_E=\frac{4GM}{b}$ is the Einstein radius of the gravitating object.
This approximation is good enough even for small $E/m$ (see Figure~\ref{function}).

It is convenient to define a relative deviation
\begin{equation}
\Delta \theta\equiv \frac{\theta-\theta_E}{\theta_E}.
\end{equation}

\section{Results}
From equation~(\ref{dispersion2}) and~(\ref{deflection}), it can be seen
that the deflection angle of a photon in a gravitational field
depends on its energy (or frequency). The relation between the
relative deviation of deflection angle from the Einstein radius, $\Delta\theta\equiv (\theta-\theta_E)/\theta_E$, 
and the ratio, $E/m$, is shown in Figure~\ref{function}. The relative deviation can be
calculated by using equation~(\ref{dispersion2}) and~(\ref{deflection}).

\begin{figurehere}
\begin{center}
\includegraphics[width=6cm]{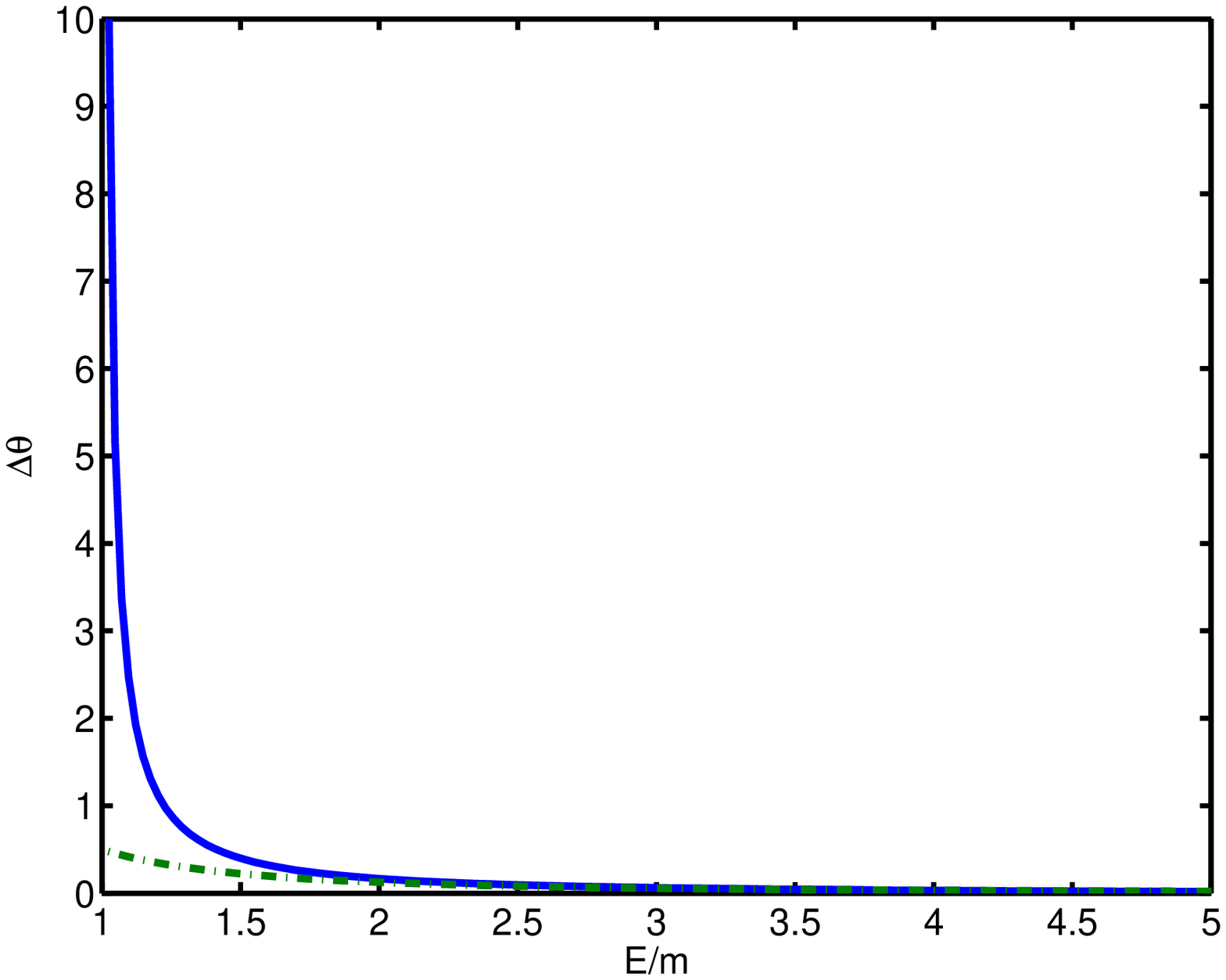}
\end{center}
\caption{\it The solid line denotes the relation between relative
deviation and the energy mass ratio $E/m$ corresponding to
equation(\ref{dispersion2}). The dot-dashed line denotes the
relation corresponding to equation(\ref{deflection}). When $E/m$,
equation (\ref{dispersion2}) is a good approximation of
equation(\ref{deflection}).}
\label{function}
\end{figurehere}

Since the relative deviation $\Delta\theta$ is energy dependent, photon mass can be constrained by comparing the relative deviation
in two different energy bands (say, $\Delta\theta_1$ at $E_1$ and
$\Delta\theta_2$ at $E_2$ ). If $E_2\gg E_1$, which means $\Delta\theta_2\ll \Delta\theta_1$, 
we can use the approximation $\Delta\theta_1\approx
\Delta\theta_1-\Delta\theta_2\approx (\theta_1-\theta_2)/\theta_2$,
where we have approximated $\theta_E$ with $\theta_2$. That is, we
can measure the deviation by comparing the gravitational lensing images
in two different energy bands.

We make use of the data from CASTLES Survey database\footnote{C.S.
Kochanek, E.E. Falco, C. Impey, J. Lehar, B. McLeod, H.-W. Rix,
http://www.cfa.harvard.edu/glensdata/}, which provides the
properties of several strong gravitational sources. The sources with
both infrared and radio data are selected and the image separations
are listed in Table 1. The first column is the name of image pairs
(e.g. PMN0134-0931 A-B means the image A and B in the source
PMN0134-0931). The second and third column are the difference of right
ascension and difference of declination between the two images of a
pair in the infrared band, while the fourth and fifth column are those in
the radio band. We denote the IR image separation with $\phi_2$ and the radio
image separation with $\phi_1$ . Since the lensing patterns at
different bands are geometrically similar,
$\phi_1/\theta_1=\phi_2/\theta_2$ , and we have
\begin{equation}
\Delta\theta=\frac{\theta_1-\theta_2}{\theta_2}=\frac{\phi_1-\phi_2}{\phi_2}.
\end{equation}
Using equation~(\ref{deflection}), $m^2=2E^2\Delta\theta$ can be
calculated, which scatter around 0 (see Figure~\ref{distribution}). The standard
deviation is  $7.59\times 10^{-77} \rm g^2$, which corresponds to an
upper limit of $8.71\times 10^{-39}\rm g$ on photon mass.

As can be seen from Figure~\ref{function}, equation~(\ref{deflection}) is a good
approximation of equation~(\ref{dispersion2}) when
$\Delta\theta^{-1}=2E^2/m^2>8$. The result above is consistent with
this condition.


\begin{figurehere}
\begin{center}
\includegraphics[width=6cm]{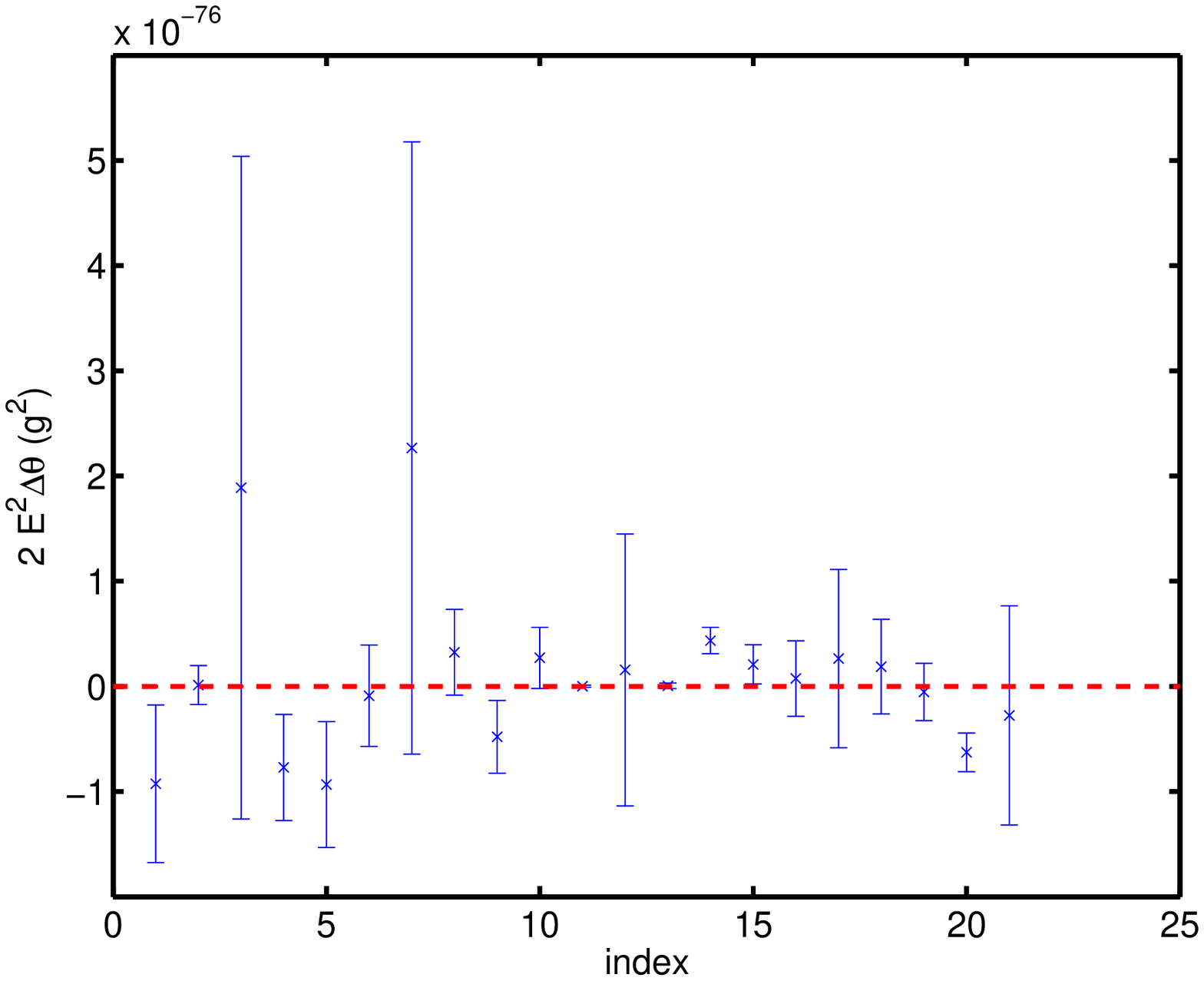}
\end{center}
\caption{\it Distribution of $2E^2\Delta\theta$. The abscissa is the index of image pairs. The horizontal line is the line corresponds to $2E^2\Delta\theta=0$.}
\label{distribution}
\end{figurehere}


\section{Discussion}
In this paper, strong gravitational lensing data are used to infer
an upper limit of photon mass, $8.71\times 10^{-39}\rm g$.

As shown in Figure~\ref{distribution}, the $2E^2\Delta\theta$ values inferred from the
lensing data are roughly consistent with 0, but with large error
bars. There are several reasons. The uncertainty in the
determination of image position may be the dominant factor.

Since $m^2=2E^2\Delta\theta$, more
accurate observation with lower frequencies can help to improve this
measurement. To improve the accuracy, longer VLBI (Very Long Baseline Interferometry) base line is needed, which is possible
with space VLBI (e.g. VSOP project in Japan). However, the improvement may be limited.

As mentioned before, in the constraints from gamma-ray burst data,
non-simultaneous emissions of photons of different energies will
affect the final result. In the current study, there are also
problems. In a quasar, radiation in different energy bands may come
from different regions. This may affect the constraint on photon
mass. This error is intrinsic and will exist no matter how
accurate the instruments are.


\Acknowledgements{\bahao The author would like to thank Dr. YUE Youling for helpful discussions.}


\normalsize \vskip0.3in\parskip=0mm \baselineskip 18pt
\renewcommand{\baselinestretch}{1.1}\footnotesize\parindent=4mm\bahao

\REF{1\ } Lakes, R. Experimental Limits on the Photon Mass and Cosmic Magnetic Vector Potential.
          Physical Review Letters, 1998, 80: 1826 Astron Astrophys, 2009, 508: L27--L30

\REF{2\ } Ellis, J., et al., Quantum-gravity analysis of gamma-ray bursts using wavelets.
          A\&A, 2003, 402: 409-424

\REF{3\ } Luo, J., et al., New Experimental Limit on the Photon Rest Mass with a Rotating
          Torsion Balance. Physical Review Letters, 2003, 90: 081801

\REF{4\ } Accioly, A. and R. Paszko, Photon mass and gravitational
          deflection. Physical Review D, 2004, 69: 107501

\REF{5\ } Lebach, D.E., et al., Measurement of the Solar Gravitational Deflection
          of Radio Waves Using Very-Long-Baseline Interferometry. Physical Review Letters, 1995, 75: 1439

\REF{6\ } Shapiro, S.S., et al., Measurement of the Solar
          Gravitational Deflection of Radio  Waves using Geodetic
          Very-Long-Baseline Interferometry Data, 1979-1999. Physical Review
          Letters, 2004, 92(12): p. 121101

\REF{7\ } Goldhaber, A.S. and M.M. Nieto, Photon and graviton mass
          limits. Reviews of Modern Physics,2010, 82: 939

\REF{8\ } Winn, J.N. et al., PMN J0134-931: A Gravitationally Lensed
          Quasar with Unusual Radio Morphology. The Astrophysical Journal,
          2002, 564: 143

\REF{9\ } Patnaik, A.R., R.W. Porcas, and I.W.A. Browne, VLBA
          observations of the gravitational lens system B0218+357. MNRAS,
          1995, 274: L5

\REF{10\ } Katz, C.A. and J.N. Hewitt, Further radio investigations
           of gravitational lensing in MG 0414+0534. The Astrophysical Journal,
           1993, 409: L9

\REF{11\ } Jackson, N., et al., B0712+472: a new radio four-image
           gravitational lens. Monthly Notices of the Royal Astronomical
           Society, 1998, 296: 483-490

\REF{12\ } Marlow, D.R. and et al., CLASS B0739+366: A New Two-Image
           Gravitational Lens System. The Astronomical Journal, 2001, 121:
           619

\REF{13\ } Lacy, M. and et al., The Reddest Quasars. II. A
           Gravitationally Lensed FeLoBAL Quasar. The Astronomical Journal,
           2002, 123: 2925

\REF{14\ } Myers, S.T. and et al., CLASS B1152+199 and B1359+154:
           Two New Gravitational Lens Systems Discovered in the Cosmic Lens
           All-Sky Survey. The Astronomical Journal, 1999, 117: 2565

\REF{15\ } Rusin, D. and et al., B1359+154: A Six-Image Lens
           Produced by a z 1 Compact Group of Galaxies. The Astrophysical
           Journal, 2001, 557: 594

\REF{16\ } Patnaik, A.R., et al., B1422+231 - A new gravitationally
           lensed system at Z = 3.62. MNRAS, 1992, 259: 1.

\REF{17\ } Marlow, D.R. and et al., CLASS B1555+375: A New
           Four-Image Gravitational Lens System. The Astronomical Journal,
           1999, 118: p. 654
\end{multicols}

\renewcommand{\baselinestretch}{1.08} \baselineskip 12.2pt\parindent=10.8pt
\begin{table}
\caption{Separation of lensing images in different bands from CASTLE
database.}
\begin{center}
\begin{tabular}{|c|c|c|c|c|c|c|}
\hline \hline \multirow{2}{*}{Image pairs} & \multicolumn{2}{|c|}{
Separation of IR image} & \multicolumn{2}{|c|}{Separation of Radio
image} & \multirow{2}{*}{Relative separation ($\Delta\theta$)} \\
\cline{2-5}
 & RA(arcsec) & Dec(arcsec) & RA(arcsec) & Dec(arcsec) &  \\
\hline
PMNJ0134-0931[8] A-B & 0.082$\pm$0.003 & 0.156$\pm$0.003 & 0.07918$\pm$0.00146 & 0.15069$\pm$0.00219 & -0.034$\pm$0.028\\
PMNJ0134-0931 A-C $^a$ & 0.539$\pm$0.003 & 0.415$\pm$0.003 & 0.53962$\pm$0.00120 & 0.41471$\pm$0.00151 & 0.000462$\pm$0.0069\\
B0218+357[9] A-B $^b$ & 0.307$\pm$0.003 & 0.126$\pm$0.003 & 0.30920$\pm$0.00014 & 0.12740$\pm$0.00014 & 0.0077$\pm$0.013\\
MG0414+0534[10] $^c$ A1-B & 0.600$\pm$0.003 & 1.942$\pm$0.003 & 0.5876$\pm$0.0003 & 1.9341$\pm$0.0003 & -0.00550$\pm$0.0021\\
MG0414+0534 A2-B & 0.732$\pm$0.003 & 1.549$\pm$0.003 & 0.7208$\pm$0.0003 & 1.5298$\pm$0.0003 & -0.01292$\pm$0.0025\\
MG0414+0534 B-C & 1.342$\pm$0.003 & 1.650$\pm$0.003 & 1.3608$\pm$0.0003 & 1.6348$\pm$0.0003 & 0.000098$\pm$0.0020\\
B0712+472[11] A-B & 0.052$\pm$0.004 & 0.146$\pm$0.007 & 0.051$\pm$0.010 & 0.160$\pm$0.010 & 0.084$\pm$0.11\\
B0712+472 A-C & 0.808$\pm$0.005 & 0.648$\pm$0.004 & 0.806$\pm$0.010 & 0.670$\pm$0.010 & 0.0119$\pm$0.015\\
B0712+472 A-D & 1.186$\pm$0.007 & 0.463$\pm$0.004 & 1.163$\pm$0.010 & 0.460$\pm$0.010 & -0.0176$\pm$0.013\\
B0739+366[12] A-B & 0.222$\pm$0.004 & 0.485$\pm$0.004 & 0.2217$\pm$0.0001 & 0.4910$\pm$0.0001 & 0.0100$\pm$0.011\\
J1004+1229[13] A-B $^d$ & 0.267$\pm$0.003 & 1.516$\pm$0.003 & 0.2633$\pm$0.0010 & 1.5172$\pm$0.0017 & 0.000354$\pm$0.0030\\
B1152+200[14] A-B $^b$ & 0.936$\pm$0.003 & 1.246$\pm$0.003 & 0.935$\pm$0.005 & 1.248$\pm$0.005 & 0.000641$\pm$0.0052\\
B1359+154[15] d A-B & 0.483$\pm$0.007 & 1.253$\pm$0.009 & 0.49020$\pm$0.00003 & 1.25240$\pm$0.00003 & 0.00152$\pm$0.0085\\
B1359+154 A-C & 0.323$\pm$0.007 & 1.640$\pm$0.003 & 0.31126$\pm$0.00003 & 1.66956$\pm$0.00003 & 0.1604$\pm$0.0046\\
B1359+154 A-D & 0.957$\pm$0.008 & 1.357$\pm$0.008 & 0.96257$\pm$0.00003 & 1.36864$\pm$0.00003 & 0.00766$\pm$0.0069\\
B1359+154 A-E & 0.627$\pm$0.013 & 1.129$\pm$0.011 & 0.60876$\pm$0.00003 & 1.14296$\pm$0.00003 & 0.00275$\pm$0.013\\
B1359+154 A-F & 0.426$\pm$0.016 & 0.951$\pm$0.028 & 0.42220$\pm$0.00003 & 0.96377$\pm$0.00003 & 0.0097$\pm$0.031\\
B1422+231[16] A-B & 0.385$\pm$0.003 & 0.317$\pm$0.003 & 0.387$\pm$0.005 & 0.320$\pm$0.005 & 0.0069$\pm$0.017\\
B1422+231 B-C & 0.336$\pm$0.003 & 0.750$\pm$0.003 & 0.332$\pm$0.005 & 0.750$\pm$0.005 & -0.00198$\pm$0.010\\
B1422+231 B-D & 0.984$\pm$0.004 & 0.802$\pm$0.003 & 0.939$\pm$0.005 & 0.810$\pm$0.005 & -0.0231$\pm$0.068\\
B1555+375[17] A-C & 0.417$\pm$0.014 & 0.013$\pm$0.008 & 0.412$\pm$0.001 & 0.028$\pm$0.001 & -0.0102$\pm$0.038\\
\hline
\end{tabular}
\end{center}
\tiny
a. {The IR image C corresponds to D in Radio observation.}

b. {The radio observation is at a frequency of 15GHz.}

c. {The radio observation is at a frequency of 8GHz.}

d. {The radio observation is at a frequency of 1.7GHz.}
\end{table}

\end{document}